\documentclass[prb,amsfonts,amssymb,floats,twocolumn,superscriptaddress,aps]{revtex4}

\usepackage[final,dvips]{epsfig}
\usepackage{amsmath}

\newcommand{\w}{\omega}
\newcommand{\llangle}{\langle\!\langle}
\newcommand{\rrangle}{\rangle\!\rangle}

\begin{document}

\title[]
{
The mass-flow error in the Numerical Renormalization Group method \\
and the critical behavior of the sub-ohmic spin-boson model
}
\author{Matthias Vojta}
\affiliation{Institut f\"ur Theoretische Physik,
Universit\"at zu K\"oln, Z\"ulpicher Str. 77, 50937 K\"oln, Germany}
\affiliation{Centro Atomico Bariloche, 8400 San Carlos de Bariloche, Argentina}
\author{Ralf Bulla}
\affiliation{Institut f\"ur Theoretische Physik,
Universit\"at zu K\"oln, Z\"ulpicher Str. 77, 50937 K\"oln, Germany}
\author{Fabian G\"uttge}
\author{Frithjof Anders}
\affiliation{Lehrstuhl f\"ur Theoretische Physik II, Technische Universit\"at Dortmund,
Otto-Hahn-Str. 4, 44221 Dortmund, Germany}

\date{\today}

\begin{abstract}
We discuss a particular source of error in the Numerical Renormalization Group (NRG)
method for quantum impurity problems, which is related to a renormalization of impurity
parameters due to the bath propagator. 
At any step of the NRG calculation, this renormalization is only partially taken into account,
leading to systematic variation of the impurity parameters along the flow.
This effect can cause qualitatively incorrect results when studying quantum critical phenomena,
as it leads to an implicit variation of the phase transition's control parameter as
function of the temperature and thus to an unphysical temperature dependence of the
order-parameter mass.
We demonstrate the mass-flow effect for bosonic impurity models with a power law bath spectrum,
$J(\w)\propto\w^s$, namely the dissipative harmonic oscillator and the spin-boson model.
We propose an extension of the NRG to correct the mass-flow error.
Using this, we find unambiguous signatures of a Gaussian critical fixed point in the
spin-boson model for $s<1/2$,
consistent with mean-field behavior as expected from quantum-to-classical mapping.
\end{abstract}


\maketitle


\section{Introduction}

The Numerical Renormalization Group method,\cite{wilson75,nrgrev}
originally developed by Wilson\cite{wilson75} for the Kondo model,
is by now an established technique for the solution of general quantum impurity problems.
It has been applied, e.g., to magnetic atoms in metals, to quantum dots and magnetic molecules,
and as an impurity solver within dynamical mean-field theory.
Its generalization\cite{BTV,BLTV} to bosonic baths has enabled the treatment of dissipative
impurity models and those with both bosonic and fermionic baths.\cite{kevin}
Quite often, impurity quantum phase transitions\cite{mvrev} are in the focus of interest.
The strengths of NRG in treating such critical phenomena lie in its ability to treat
arbitrarily small energy scales and in its renormalization-group character
which allows e.g. for the analysis of flow diagrams.

Recently, conflicting results have been reported about the critical behavior of certain impurity
models with a bosonic bath, in particular the spin-boson and the Ising-symmetric Bose-Fermi Kondo
model.\cite{bosonization_foot}
For a bosonic bath with power-law spectral density $J(\w)\propto\w^s$, these models
display a quantum phase transition for $0<s\leq1$.
Statistical-mechanics arguments suggest that this transition is in the same universality class
as the thermal phase transition of the one-dimensional (1d) Ising model with $1/r^{1+s}$
long-range interactions.
At issue is the validity of this quantum-to-classical correspondence for $s<1/2$
where the Ising model is above its upper-critical dimension and displays mean-field
behavior.\cite{luijten,fisher}
Initially, two of us claimed non-classical behavior with hyperscaling
in the spin-boson model for $s<1/2$,
based primarily on NRG results.\cite{VTB}
These results have been verified by others,\cite{karyn} and extended
to the Ising-symmetric Bose-Fermi Kondo model.\cite{kevin}
In contrast, subsequent Quantum Monte Carlo (QMC)\cite{rieger,werner} and
exact-diagonalization\cite{fehske} studies  concluded that the critical behavior
of the spin-boson model for $s<1/2$ is classical and of mean-field type.
We have recently retracted the claim\cite{VTB} of non-classical behavior, because we have
realized two different sources of error of the NRG which spoil the determination of
critical exponents.\cite{erratum}
However, other authors continue to rely on NRG results in this context.\cite{si09,glossop09}

\begin{figure}[!b]
\includegraphics[width=2.5in,clip]{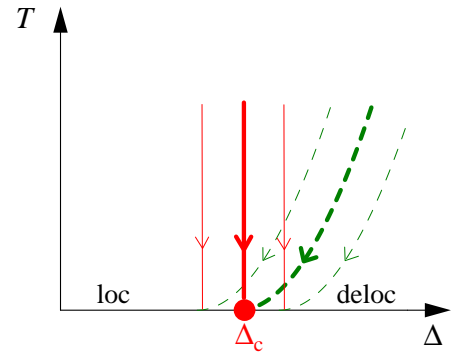}
\caption{
Schematic phase diagram of the spin-boson model $\mathcal{H}_{\rm SB}$,
as function of temperature $T$ and tunneling strength $\Delta$
(keeping the dissipation strength $\alpha$ fixed),
showing a path with $\Delta\!=\!\Delta_c$ (thick solid) along which quantum-critical observables
should be measured.
Due to the mass-flow effect, a temperature-dependent deviation of the
order-parameter mass is induced, such that a system located at $\Delta_c$ at $T\!=\!0$
follows a path with finite mass, $\delta\Delta\propto T^s$, at any $T>0$ (thick dashed).
The thin lines represent trajectories
with (dashed) and without (solid) mass-flow effect for $\Delta \neq \Delta_c$.
}
\label{fig:pdflow}
\end{figure}

In this paper, we investigate one of the error sources of the NRG in more detail,
which we have dubbed the mass-flow effect. It arises from the NRG algorithm which iteratively
integrates out the impurity's bath.
For a particle-hole asymmetric bath, the real part of the bath propagator generates
a physical shift of impurity parameters;
for models with single-particle tunneling between impurity and bath it is simply the energy
of the impurity level which is shifted due to the real part of the hybridization
function.
As a NRG calculation ignores the part of the bath spectrum below the current NRG scale,
there is, at any NRG step, a {\em missing} parameter shift which is is set by the
current NRG scale.
Near a quantum phase transition, this implies an artificial scale-dependent shift of the
order-parameter mass.
For a NRG calculation with model parameter values corresponding to the critical point,
the system is therefore {\em not} located at the critical coupling for any finite $T$,
but effectively follows a trajectory in the phase diagram as sketched in
Fig.~\ref{fig:pdflow}.
This spoils the measurement of critical properties extracted in a NRG run as function
of $T$.

Other sources of error within the NRG method are the discretization of the continuous
bath density of states, the truncation of the eigenvalue spectrum in each NRG step to the
lowest $N_s$ states, and the truncation of the bath Hilbert space in the case of a
bosonic bath, where only $N_b$ states are taken into account.
While the effects of discretization and spectrum truncation are well studied and understood within the
fermionic NRG,\cite{wilson75,nrgrev} Hilbert-space truncation is more serious.
Ref.~\onlinecite{BLTV} pointed out that it precludes a correct representation of the ordered
phase of the spin-boson model for $s<1$ at low energies or temperatures. Later, it was
realized\cite{erratum} that it also leads to incorrect results for the order-parameter
exponents $\beta$ and $\delta$ of the phase transition above the upper-critical
dimension.
In this paper, our focus will be on the mass-flow effect; the other errors will be discussed
when appropriate.

First, we shall demonstrate the mass-flow effect within a model of non-interacting bosons,
namely the dissipative harmonic oscillator, for which all statements can be made exact.
In this model, the critical point translates into the instability point where the renormalized
impurity energy is zero.
The mass flow will be shown to lead to qualitatively incorrect results; this problem
carries over to interacting models, like the anharmonic oscillator or the spin-boson
model, if the critical point is Gaussian (i.e. above its upper-critical dimension).
Second, we propose an extension of the iterative diagonalization scheme to cure the
mass-flow error. This extension solves the problem for the full parameter range of
the non-interacting harmonic oscillator, while working asymptotically for models of
interacting bosons.
Third, we apply the extended NRG algorithm to the spin-boson model.
For $s<1/2$, we find results qualitatively different from those\cite{BTV,VTB} of the standard
NRG implementation: our new results signify a flow towards a Gaussian critical fixed
point.
While the truncation of the bosonic Hilbert space precludes calculations very close to
this Gaussian fixed point, we can identify a mean-field power law in the impurity susceptibility.
Taken together, this shows that -- as other methods -- also the NRG predicts that
the spin-boson model exhibits mean-field behavior for $s<1/2$.

It is worth noting that an observation reminiscent of the mass-flow effect has been made
in Ref.~\onlinecite{si09}: the critical behavior of the classical long-range Ising
model with $s<1/2$ was found to change from mean-field-like to hyperscaling-like upon
artificially truncating the ``winding'' of the long-range interaction (i.e. upon violating the
periodic boundary conditions in imaginary time). This finding underscores that mean-field
critical behavior in the long-range models under consideration can be easily spoiled
by algorithmic errors. Note, however, that we disagree with the interpretation regarding
the quantum-to-classical correspondence given in Ref.~\onlinecite{si09}, see below.

\subsection{Outline}

The remainder of the paper is organized as follows:
In Sec.~\ref{sec:models} the model Hamiltonians are introduced.
Sec.~\ref{sec:chain} explains how the mass-flow effect arises from the iterative
diagonalization of the Wilson chain. The dissipative harmonic oscillator is subject of
Sec.~\ref{sec:dho}, where the mass-flow error in the susceptibility is demonstrated
analytically. This knowledge is used in Sec.~\ref{sec:cure} to propose a modification of
the iterative-diagonalization scheme, designed to cure the mass-flow error.
Finally, the modified NRG algorithm is applied to the spin-boson model in
Sec.~\ref{sec:sb}. The NRG flow is discussed separately for $s>1/2$ and $s<1/2$ and
compared to the results from standard NRG.
The results are interpreted in terms of a Gaussian critical fixed point for $s<1/2$.
Conclusions close the paper.
Various details, including a discussion of the mass-flow effect in fermionic impurity models
with particle--hole asymmetry, are relegated to the appendices.


\section{Models}
\label{sec:models}

The mass-flow effect can be most easily demonstrated using impurity models of
non-interacting particles.
We shall consider the dissipative harmonic oscillator, with the Hamiltonian
\begin{eqnarray}
{\cal H}_{\rm DHO} &=& \Omega a^\dagger a + \frac{\epsilon}{2}(a+a^\dagger) \nonumber\\
&+& \frac 1 2 \sum_{i} \lambda_{i} (a + a^\dagger) ( b_{i} + b_{i}^{\dagger} ) +
\sum_{i} \w_{i} b_{i}^{\dagger} b_{i} ,
\label{dho}
\end{eqnarray}
where $\Omega>0$ is the bare ``impurity'' oscillator frequency, $\epsilon$ is a field
conjugate to the oscillator position,
and the $\w_i>0$ are the frequencies of the bath oscillators.
The bath is completely specified by its propagator at the ``impurity'' location
\begin{equation}
\label{gw}
    \Gamma(\w)= \sum_{i} \frac{\lambda_{i}^2}{\omega+i0^+ -\omega_{i}}
\end{equation}
with the spectral density
\begin{equation}
\label{jw}
    J(\w)= - {\rm Im}\,\Gamma(\w) = \pi \sum_{i} \lambda_{i}^{2} \delta\left( \omega -\omega_{i}\right)\,.
\end{equation}

Universal properties of impurity phase transitions are determined
by the behavior of the low-energy part of the bath spectrum $J(\w)$.
Discarding high-energy details, the common parametrization is
\begin{equation}
  J(\omega) = 2\pi\, \alpha\, \omega_c^{1-s} \, \omega^s\,,~ 0<\omega<\omega_c\,,\ \ \ s>-1
\label{power}
\end{equation}
where the dimensionless parameter $\alpha$ characterizes the
dissipation strength, and $\omega_c$ is a cutoff energy.
The value $s=1$ represents the case of ohmic dissipation.

The dissipative oscillator with a power-law bath spectrum is known to become unstable at
large dissipation:\cite{weiss}
The coupling to the bath renormalizes the oscillator frequency
$\Omega$ downwards, which becomes zero at some $\alpha_c$.
Hence, the behavior of the model is not well-defined for $\alpha>\alpha_c$.

The system at large dissipation may be stabilized by adding a local repulsive interaction
to ${\cal H}_{\rm DHO}$. A symmetry-broken phase can emerge, with ``condensation'' of the $a$
bosons.
Two possible routes are
\begin{equation}
\label{dao}
{\cal H}_{\rm DAO}= {\cal H}_{\rm DHO} + U n_a (n_a-1),~~n_a = a^\dagger a
\end{equation}
and
\begin{equation}
\label{dao2}
{\cal H}'_{\rm DAO}= {\cal H}_{\rm DHO} + u (a+a^\dagger)^4.
\end{equation}
The latter, ${\cal H}'_{\rm DAO}$, can be understood as a local $\phi^4$ impurity.
On the other hand, ${\cal H}_{\rm DAO}$ in the limit $U=\infty$ becomes equivalent to
the standard spin-boson model
\begin{equation}
{\cal H}_{\rm SB}=-\frac{\Omega}{2}\sigma_{x} + \frac{\epsilon}{2}\sigma_{z}
+\frac{\sigma_{z}}{2} \sum_{i}
    \lambda_{i}( b_{i} + b_{i}^{\dagger} )
+\sum_{i} \omega_{i} b_{i}^{\dagger} b_{i}
\label{sbm}
\end{equation}
where $\sigma_z=\pm 1$ are the local impurity states, and $\Omega$ is the tunneling rate.
The equivalence is seen by identifying the remaining oscillator states $|0\rangle$ and $|1\rangle$
in ${\cal H}_{\rm DAO}$ with the states $(|\uparrow\rangle\pm|\downarrow\rangle)/\sqrt{2}$.
In all three models (\ref{dao},\ref{dao2},\ref{sbm}),
the ordered phase at large dissipation breaks an Ising symmetry,
$a\leftrightarrow -a$ (or $\sigma_z\leftrightarrow -\sigma_z$), $b\leftrightarrow -b$,
and is associated with a non-zero expectation value $\langle a+a^\dagger\rangle$
(or $\langle\sigma_z\rangle$).

Universality arguments suggest that the critical properties of the phase transitions
are identical in the three models and coincide with those of a classical Ising chain
with $1/r^{1+s}$ interactions.
This quantum-to-classical correspondence trivially holds for
${\cal H}'_{\rm DAO}$ in Eq.~\eqref{dao2}, as its imaginary-time path integral
representation at $T\!=\!0$ is identical to the continuum limit of the
one-dimensional Ising model (i.e. a scalar $\phi^4$ theory).\cite{fisher}
For $s<1/2$, the critical behavior is Gaussian and mean-field like,
with the quartic interaction being dangerously irrelevant at criticality.

The quantum phase transition in the spin-boson model has been extensively studied:
While the ohmic case, $s=1$, has long been known to display a Kosterlitz-Thouless
transition,\cite{leggett} the sub-ohmic case has only been investigated more
recently.\cite{KM96,BTV,VTB}
For $0<s<1$, a continuous quantum phase transition emerges, with critical exponents
depending on $s$.
While there is consensus that, for $1/2<s<1$, those exponents are identical to the ones
of the corresponding 1d Ising model,
a debate is centered around the issue of whether or not this continues to hold for $0<s<1/2$
where the Ising model displays mean-field behavior.
Alternatively, non-mean-field exponents obeying hyperscaling have been proposed on the
basis of NRG calculations\cite{VTB,karyn}, and also carried over to the Ising-symmetric
Bose-Fermi Kondo model.\cite{bosonization_foot,kevin,si09}
In particular, NRG has been used to calculate the local susceptibility $\chi$ at
the critical coupling as function of temperature, which was found to follow a power
law $\chi\propto T^{-x}$ with $x=s$.
In contrast, mean-field behavior
implies\cite{luijten} $x=1/2$, which has indeed been found e.g. using QMC simulations.
\cite{rieger}

We shall argue here, expanding on our previous note,\cite{erratum}
that the proposals of non-classical behavior are erroneous for the spin-boson model
and questionable for the Ising-symmetric Bose-Fermi Kondo model.
For the former, we show that the critical behavior instead is of
mean-field type, consistent with numerical studies using
QMC and exact-diagonalization methods.\cite{rieger,werner,fehske}


\section{Wilson chain and mass flow}
\label{sec:chain}

\begin{figure}[!b]
\includegraphics[width=2.9in,clip]{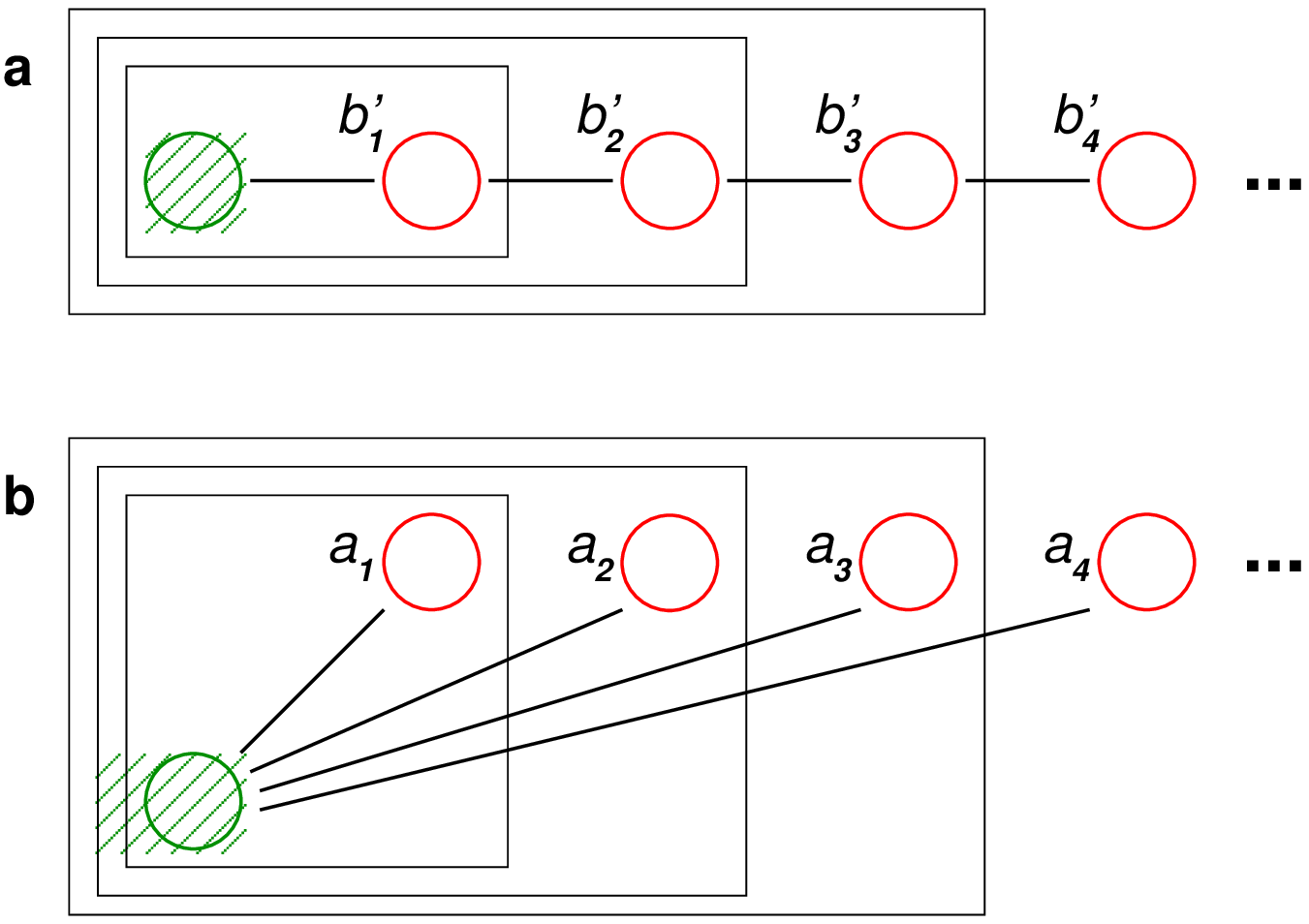}
\caption{
Structure of the NRG Hamiltonian, with the bath represented by a semi-infinite Wilson
chain, with bath operators $b'_n$. The boxes indicate the iterative diagonalization
scheme.
}
\label{fig:itdiag}
\end{figure}

Within the NRG algorithm, the bath is represented by a semi-infinite (``Wilson'') chain,
Fig.~\ref{fig:itdiag},
such that the local density of states at the first site of this chain is a discrete
approximation to the bath density of states.\cite{wilson75,nrgrev}
Due to the logarithmic discretization, the site energies $\epsilon_n$ and hopping matrix
elements $t_n$ decay exponentially along the chain according to $\w_c \Lambda^{-n+1}$,
where $\Lambda$ is the discretization parameter.

Let us denote by $\mathcal{H}_n$ the Hamiltonian of impurity plus $n$ sites of the Wilson
chain, and by $\Gamma_n(\w)$ the propagator at the impurity site of this $n$-site bath.
Then, $\mathcal{H}_\infty$ is the discretized version of the original problem.
During the NRG run, $\mathcal{H}_\infty$ is diagonalized iteratively: First,
$\mathcal{H}_1$ is diagonalized and the lowest $N_s$ eigenstates are kept. Then, the next
bath site is added to form $\mathcal{H}_2$, the new system is diagonalized, and again the
lowest $N_s$ eigenstates are kept (which are approximations to the lowest states of
$\mathcal{H}_2$). As the characteristic energy scale of the low-lying part of the
eigenvalue spectrum decreases by a factor of $\Lambda$ in each step, this process is
repeated until the desired lowest energy is reached.
Temperature-dependent thermodynamic observables at a temperature $T_n = \w_c \Lambda^{-n+1} / \bar\beta$
are typically calculated via a thermal average taken from the eigenstates at NRG step $n$.
Here, $\bar\beta$ is a parameter of order unity which is often chosen as
$\bar\beta=1$.

The iterative diagonalization procedure implies that, at NRG step $n$, the chain sites $n+1$,
$n+2$, \ldots have not yet been taken into account, i.e., the effect of those sites does
not enter thermodynamic observables at temperature $T_n$.
Typically, this is a reasonable approximation,
as the spectral density of the missing part of the chain,
${\rm Im}(\Gamma_\infty-\Gamma_n)(\w)$,
has contributions at energies below $\w_c \Lambda^{-n}$ only.

However, the missing chain also implies a missing contribution to the real part of
the bath propagator.
This can be easily estimated:
For a power-law bath spectrum, Eq.~\ref{power}, the zero-frequency real part
${\rm Re}(\Gamma_\infty-\Gamma_n)(\w\!=\!0)$ is generated by frequencies
$0<\w<\w_c\Lambda^{-n}$ and scales as $\w_c\Lambda^{-ns}$, i.e.,
up to numerical factors it scales as the NRG energy scale $T_n$ to the power $s$.
As we will show below, this missing real part implies a flow of the order-parameter mass
and can spoil the analysis of critical phenomena.

To support the above estimate, we calculate the local Green's function $G_n^0$
at the initial site, $b'_1$, of the Wilson chain (which is proportional to $\Gamma_n(\w)$) for
different chain lengths $n$.
To this end, we numerically diagonalize the single-particle problem corresponding to a
Wilson chain with parameters $\epsilon_n$ and $t_n$ chosen to represent a power-law bath spectrum,
Eq.~\eqref{power}, as in the NRG.\cite{nrg_param}

\begin{figure}[!t]
\epsfxsize=3.5in
\centerline{\epsffile{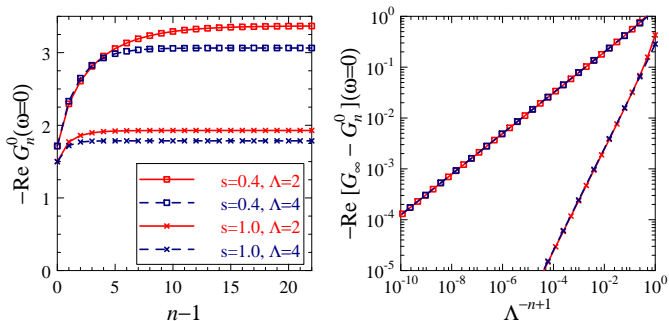}}
\caption{
a) Zero-frequency real part of the bath propagator at the initial site of the Wilson chain,
$-{\rm Re\,}G_n^0(\w=0)$, as function of the chain length $n$, for different
bath exponents $s$ and discretization parameters $\Lambda$, all with $\w_c=1$.
b) The piece of the real part which is missing at chain length $n$,
$-{\rm Re\,}(G_\infty^0 - G_n^0)$, for the same parameters,
plotted as function of the characteristic energy scale of
the Wilson chain, $\Lambda^{-n+1}$.
}
\label{fig:reg}
\end{figure}

Explicit results for ${\rm Re\,}G_n^0(\w=0)$ are shown in Fig.~\ref{fig:reg}a.
As expected, ${\rm Re\,}G_n^0$ approaches a finite (negative) value as $n\to\infty$,
which depends on both $s$ and $\Lambda$.
The missing real part ${\rm Re\,}(G_\infty^0 - G_n^0)$ is shown in panel b
and scales as $T_n^s$, with a prefactor which depends on $s$, but only weakly on
$\Lambda$.

\section{Dissipative harmonic oscillator}
\label{sec:dho}

We shall discuss how the mass-flow effect influences observables for the
simplest model, the dissipative harmonic oscillator \eqref{dho}.
It is important to distinguish the various methods to calculate observables
in this non-interacting model:
(i) For a continuous power-law spectrum, a number of quantities can be calculated
analytically.
(ii) For a discretized bath, represented by a semi-infinite Wilson chain, the
single-particle problem can be solved by exact diagonalization for long chains.
(iii) As in NRG, one may use a truncated Wilson chain with temperature-dependent
length, and again diagonalize the single-particle problem.
(iv) A true NRG calculation can be performed, which treats the full many-body problem.
Here, we shall mainly be interested in comparing the results of (ii) and (iii), which
allows to assess the mass-flow error.
In contrast, the difference between (i) and (ii) can be used to quantify the discretization
error, while the difference between (iii) and (iv) is due to spectrum and Hilbert-space truncation
of NRG.

The most interesting observable is the susceptibility associated with the oscillator
position, defined according to
\begin{equation}
\chi = d\langle a+ a^\dagger\rangle / d\epsilon,
\label{chi_ho}
\end{equation}
which is the analogue of $d\langle\sigma_z\rangle/d\epsilon$ in
the spin-boson model.
Importantly, $\chi$ is given by a single-particle propagator, $\chi = -G_x$,
with
\begin{eqnarray}
G_x(\w)
&=& \llangle a+a^\dagger;a+a^\dagger \rrangle \\
&=& \frac {2\Omega}{\w^2 + i0^+ -\Omega^2 - \Omega [\Gamma(\w) + \Gamma(-\w)] / 2},
\nonumber
\end{eqnarray}
note the factors of $1/2$ in Eq.~\eqref{dho}.
This equation shows that the dissipative oscillator is unstable at and beyond the ``resonance''
which occurs at some dissipation strength $\alpha_c$, defined
by $\Omega + {\rm Re} \Gamma(\w\!=\!0)=0$.
For $\alpha<\alpha_c$, all eigenenergies of the system are positive,
whereas the lowest one turns negative for $\alpha>\alpha_c$.
Thus, $\alpha_c$ corresponds to a singularity of the dissipative harmonic oscillator,
separating the stable from the unstable regime.


\begin{figure}[!t]
\epsfxsize=3.1in
\centerline{\epsffile{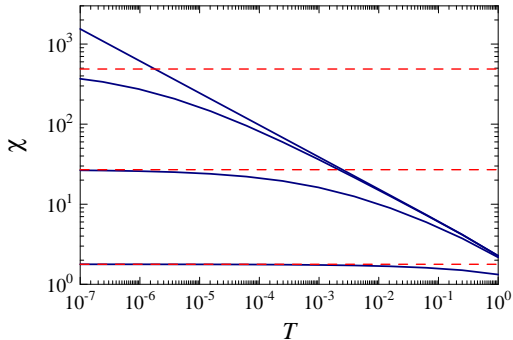}}
\caption{
Mass-flow error for the dissipative harmonic oscillator:
the graph shows the susceptibility $\chi$, Eq.~\eqref{chi_ho},
calculated at temperature $T_n$ by diagonalizing the single-particle problem
of a truncated $n$-site Wilson chain.
Parameters are $\Omega=1$, $s=0.4$, $\w_c = 1$, $\Lambda = 4$ where $\alpha_c= 0.2284682$.
The different curves are for $\alpha=0.1$, 0.22, 0.228, $\alpha_c$ (from bottom to top).
The dashed lines denote the (exact) temperature-independent $\chi$ for the $\alpha<\alpha_c$,
obtained from the semi-infinite Wilson chain.
As an aside, we note that the critical coupling for a continuum bath spectrum
is given by $\alpha_c = s \Omega / (2 \w_c)$. This evaluates to 0.2, showing
that the discretization error is around 15\%.
}
\label{fig:chi_ho}
\end{figure}

Returning to the susceptibility $\chi$, its static limit evaluates to
\begin{eqnarray}
\label{chistat}
\chi = \frac{2}{\Omega + {\rm Re} \Gamma(\w=0)}
\end{eqnarray}
which is seen to be temperature-independent and only determined by $\Omega$ and
the real part of the bath propagator.
Consequently, there is a strong mass-flow effect,
as the renormalized oscillator frequency in the denominator of $\chi$ reads
$\Omega_r = \Omega + {\rm Re} \Gamma_n(\w=0)$ for a $n$-site chain.
This is illustrated in Fig.~\ref{fig:chi_ho} where we show the susceptibility as function of
temperature, calculated using either a long Wilson chain for all $T$ or
an $n$-site Wilson chain at temperature $T_n$, i.e., using methods (ii) and (iii)
described above.
Most importantly, $\chi$ calculated from the truncated Wilson chain is
temperature-dependent, in contrast to the exact result.
For $\alpha<\alpha_c$, the exact result is approached at low $T$.
The error is most drastic at resonance, $\alpha=\alpha_c$.
There, $\chi_{\rm exact}=\infty$ (i.e. the system is unstable),
whereas the calculation using a truncated Wilson chain gives $\chi \propto T^{-s}$.
Physicswise, the mass-flow effect introduces a finite and temperature-dependent
oscillator frequency $\Omega_r \propto T^s$,
thus artificially stabilizing the system at $\alpha_c$.
Naturally, the same result is found using a full NRG calculation.

As the dissipative harmonic oscillator represents the fixed-point Hamiltonian of the
Gaussian critical point of, e.g., the anharmonic oscillator ${\cal H}'_{\rm DAO}$ in Eq.~\eqref{dao2},
it is straightforward to discuss the mass-flow effect there.
The renormalized $\Omega_r$ can be identified with the
order-parameter mass, and $\chi(\w=0)=1/\Omega_r$.
Along the flow towards the Gaussian fixed point, the irrelevant interaction $u$
leads to an order-parameter mass $\propto T^{1/2}$ for $s<1/2$,
and the physical susceptibility follows $\chi \propto T^{-1/2}$ (Ref.~\onlinecite{luijten}).
However, the artificial mass $\propto T^s$ caused by the mass-flow effect dominates the
physical mass at low $T$, leading again to the unphysical result $\chi \propto T^{-s}$.
As this coincides with the physical result for an interacting critical fixed point
with hyperscaling, the unphysical result from the mass-flow effect could be mistaken
as a signature of interacting quantum criticality.

We should emphasize that a renormalization-group scheme which successively integrates out
the impurity's bath is perfectly valid. However, it requires that the calculation of observables
at some scale $T$ accounts for the remaining part of the bath. The latter is not the case
in the iterative diagonalization scheme of standard NRG.


\section{Cure of mass flow}
\label{sec:cure}

The mass-flow error arises from the missing real part of the bath propagator,
${\rm Re}\Gamma_n(\w=0)$,
which, for every step of the iterative diagonalization, is simply a number.
Ideally, a general algorithmic solution of the mass-flow problem would directly
correct ${\rm Re}\Gamma_n$. However, this is limited by Kramer-Kronig relations, and
we have not found a manageable implementation of this idea.

In the following, we shall instead make use of physics arguments in order to (approximately)
correct the mass-flow error.
For the harmonic oscillator, ${\rm Re}\Gamma_n$ directly renormalizes the oscillator's energy,
while things are conceptually more complicated for interacting models (like the
spin-boson or Bose-Fermi Kondo models).
Therefore, we shall separately discuss the non-interacting and interacting cases in the
following.

\subsection{Dissipative harmonic oscillator}
\label{sec:curefree}

A simple recipe can be used to correct the mass-flow error when diagonalizing
a finite-length chain corresponding to the harmonic-oscillator $\mathcal{H}_n$.
We define a Hamiltonian piece ${\mathcal K}_n$ by
\begin{equation}
\label{kn}
{\mathcal K}_n = \mathcal{R}\,{\rm Re}(\Gamma_\infty-\Gamma_n)(\w\!=\!0)
\end{equation}
with $\mathcal{R} = a^\dagger a$.
As a result, $\mathcal{H}_n+{\mathcal K}_n$ has the correct
mass term, i.e., the correct renormalized oscillator frequency, for any $n$,
and diagonalizing $\mathcal{H}_n+{\mathcal K}_n$ instead of $\mathcal{H}_n$
in step $n$ removes the mass-flow problem.
One obtains the correct result for $\chi$:
thanks to ${\mathcal K}_n$, the denominator of $\chi$ in Eq.~\eqref{chistat}
is replaced by
$\Omega + {\rm Re}(\Gamma_\infty-\Gamma_n)(\w\!=\!0) + {\rm Re}\Gamma_n(\w\!=\!0)$
which is the exact result for the semi-infinite Wilson chain.

\subsection{NRG implementation}
\label{sec:curenrg}

A mass-flow correction via ${\mathcal K}_n$ can be straightforwardly implemented into
the iterative diagonalization scheme of the NRG method.
The modified NRG algorithm (dubbed NRG$^\ast$ in the following) works as follows:
(i) Initially, one diagonalizes $\mathcal{H}''_1 = \mathcal{H}_1+{\mathcal K}_1$.
In addition to the usual observables, the matrix elements of the operator $\mathcal R$
are stored as well.
Then, the following steps are repeated:
(ii) From the lowest $N_s$ states of the solution of NRG step $n$ and the states of
the impurity site $n+1$, one constructs $\mathcal{H}'_{n+1}$.
In contrast to $\mathcal{H}_{n+1}$, the operator $\mathcal{H}'_{n+1}$ contains a
mass-flow correction from the previous steps.
(iii) Using the matrix elements of $\mathcal{R}$, one constructs
$\mathcal{H}''_{n+1} = \mathcal{H}'_{n+1} + {\mathcal K}_{n+1}-{\mathcal K}_n$.
(iv) One diagonalizes $\mathcal{H}''_{n+1}$ and re-calculates the matrix elements of
the desired observables and of $\mathcal R$.

The correction of the mass-flow error, contained in steps (i) and (iii) which
differ from the usual NRG algorithm, is implemented such that the frequency shifts
cancel in the limit $n\to\infty$.
Hence, runs of NRG and NRG$^\ast$ with the same model parameters should target the
same point in the phase diagram as $T\to0$ (although their finite-temperature
trajectories are different, Fig.~\ref{fig:pdflow}).
However, this is only true in the absence of spectrum truncation.
For finite $N_s$, the cancellation is only approximate, i.e.,
there will be a small (but unimportant) parameter shift due to the mass-flow correction.

\subsection{Beyond non-interacting bosons}
\label{sec:cureint}

Being interested in extracting critical properties, we identify the mass-flow effect as a
scale-dependent shift of the order-parameter mass. This suggests that the mass flow can
be corrected by an appropriate shift in the phase transition's control parameter -- this
is simply a generalization of Eq.~\eqref{kn} where ${\mathcal K}_n$ shifts the oscillator
frequency.
We thus propose to employ a correction of the form
\begin{equation}
\label{kn2}
{\mathcal K}_n = \kappa\,\mathcal{R}\,{\rm Re}(\Gamma_\infty-\Gamma_n)(\w\!=\!0)
\end{equation}
where $\mathcal{R}$ is now a (local) operator which can be used to tune the phase transition,
e.g., the tunneling term $\sigma_x$ in the spin-boson model or the Kondo coupling
term in a Bose-Fermi Kondo model.
Importantly, the required shift will no longer be identical to ${\rm Re}(\Gamma_\infty-\Gamma_n)$.
This is already clear for the dissipative anharmonic oscillators, Eqs.~\eqref{dao} and
\eqref{dao2}, where the quartic interaction will renormalize both the oscillator frequency
and also its shift due to the bath, but in a different fashion.
Hence, we have introduced the non-universal prefactor $\kappa$ which we intend to determine
by physical criteria.

Two issues require special consideration:
(a) Is the linear relation between the required shift in the control parameter and the
missing real part of $\Gamma$, which is implied by Eq.~\eqref{kn2}, justified?
(b) How can one determine the prefactor $\kappa$?

The simplest argument for (a) is as follows:
The phase transition's control parameter
(equivalently, the distance to criticality or the bare order-parameter mass)
depends on both the prefactor of $\mathcal{R}$ and the real part of ${\rm Re}\Gamma(\w\!=\!0)$.
Both dependencies have a regular Taylor expansion at a given point in parameter space,
hence, the leading terms are linear.
As ${\rm Re}\Gamma$ changes by a known amount in every step of the iterative diagonalization
due to the mass-flow effect,
this can be compensated by a change in the prefactor of $\mathcal{R}$
which proportional to this amount, i.e., a change of the form
${\mathcal K}_{n+1}-{\mathcal K}_n$ with some fixed $\kappa$.
This argument only relies on the Taylor expansion and is thus asymptotically correct
for small changes in ${\rm Re}\Gamma$, i.e., for $T\to 0$.
(For a given model, like the anharmonic oscillator \eqref{dao2}, one can check the linear behavior by
an explicit perturbative calculation.)
Physically, it is clear that the linear term of the expansion will capture
the correct behavior in the vicinity of a given renormalization-group fixed point,
i.e., the required $\kappa$ depends on the fixed point of interest (and on non-universal
high-energy details).
Note, however, that the procedure is more general than these considerations
suggest: As both the Gaussian critical fixed point and the delocalized fixed point are
asymptotically non-interacting, a fixed $\kappa$ can be used to capture the entire
crossover from the quantum critical to the delocalized regime in this case.

Question (b) will be discussed for different types of critical fixed points in turn.
We shall use the language of the dissipative anharmonic oscillator \eqref{dao2},
where the critical theory is known.\cite{fisher,luijten}

\subsubsection{Gaussian critical fixed points}

A Gaussian critical fixed point, realized for $s<1/2$,
provides a simple criterion to find the correct value $\kappa_0$ of the correction
parameter $\kappa$, namely the temperature dependence of the order parameter mass.
As emphasized in Sec.~\ref{sec:dho}, the artificial mass generated by the mass-flow
effect follows $T^s$, while the physical mass scales as $T^{1/2}$.
Thus, in general the mass at the critical coupling will be given by
$\lambda_1(\kappa_0-\kappa) T^s + \lambda_2 T^{1/2}$ where $\lambda_{1,2}$ are prefactors.
For $\kappa<\kappa_0$ (undercompensation), the positive $T^s$ term will always dominate at low $T$
and mimic hyperscaling properties.
For $\kappa>\kappa_0$ (overcompensation), the mass will become negative at low $T$, i.e.,
the flow will be towards the localized phase.
An intermediate flow inside the localized phase will even occur if couplings
are chosen to be slightly in the delocalized phase: for $\Delta\gtrsim\Delta_c$ or
$\alpha\lesssim\alpha_c$ the system flows from critical to localized and then back to
delocalized upon lowering $T$, accompanied by a non-monotonic behavior of $\chi$.

This suggests the following simple recipe to determine $\kappa_0$:
Start with large $\kappa$ such that non-monotonic flows are seen near the critical coupling.
Decrease $\kappa$ until those disappear and the susceptibility follows a power law different
from hyperscaling at the critical coupling down to the lowest accessible temperatures.
If $\kappa$ is decreased too far, then $\chi\propto T^{-s}$ is recovered.
Hence, a clear signature of Gaussian criticality is a qualitatively different
behavior in $\chi$ for small and large $\kappa$.
In Sec.~\ref{sec:sb} and App.~\ref{app:kappa}, we shall demonstrate this for the
spin-boson model at $s<1/2$.

\subsubsection{Interacting critical fixed points}

In the case of an interacting critical fixed point, realized for $s>1/2$,
hyperscaling is fulfilled on physical grounds. Hence, the mass will invariably scale as
$T^s$ at the critical coupling, both for $\kappa<\kappa_0$ and $\kappa>\kappa_0$.
This simply reflects the fact that the mass-flow effect does not introduce qualitative
(but only quantitative) errors here, in contrast to the case of Gaussian criticality.
Hence, the behavior in the quantum-critical regime does not provide a sharp physical
criterion to determine $\kappa_0$.
We conclude that a clear signature of true interacting criticality is an insensitivity to
the value of $\kappa$ of the qualitative critical behavior.

For the spin-boson model at $s=1$, a comparison of observables to those from other
solutions like Bethe Ansatz or bosonization could be used to determine $\kappa_0$
(for either the localized or the delocalized phase).
As $s=1$ plays the role of a lower-critical dimension, we have not followed this route
further.


\section{Spin-boson model}
\label{sec:sb}

We now apply the modified NRG$^\ast$ algorithm, which includes the mass-flow
correction \eqref{kn2}, to the spin-boson model.
Note that we will make no a-priori assumptions on the nature of the critical
fixed points, but instead apply the strategies outlined in Sec.~\ref{sec:cureint}
to determine the optimal $\kappa_0$ within the NRG$^\ast$ algorithm.

\subsection{Flow diagrams}

We have studied the flow diagrams for various values of the bath exponent $0<s<1$ and
the mass-flow correction parameter $\kappa$.
While a detailed set of data is displayed in App.~\ref{app:kappa}, the main
conclusion is that for $s<1/2$ the flow changes qualitatively as $\kappa$ is varied,
while this is not the case for $s>1/2$. The former fact can be used to determine
$\kappa_0$ for $s<1/2$, while a rough estimate of $\kappa_0$ for $s\gtrsim 1/2$ may be obtained
from an extrapolation of $\kappa_0(s)$.

\begin{figure}[!t]
\epsfxsize=3.6in
\centerline{\epsffile{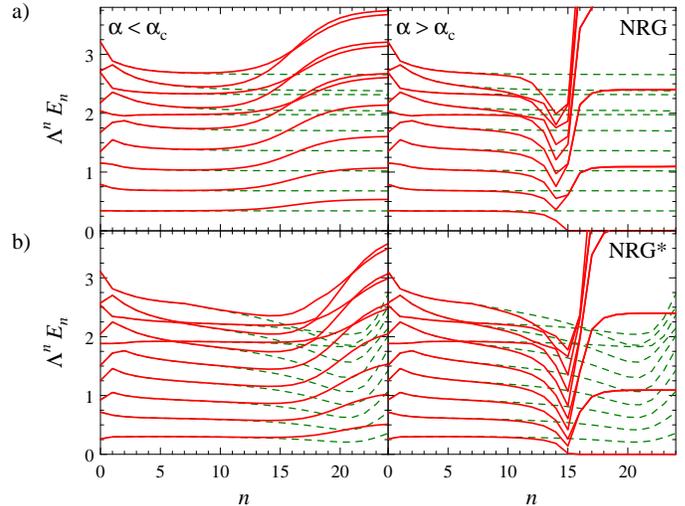}}
\caption{
NRG flow showing the 10 lowest levels
for the spin-boson model at $s=0.4$ obtained from
a) standard NRG without mass-flow correction and
b) NRG$^\ast$ with $\kappa=0.5$ mass-flow correction.
The NRG parameters are $\Lambda=4$, $N_b=12$, $N_s=40$,
while the remaining model parameters are $\Delta=\w_c=1$.
In a), $\alpha=0.3555$ (left) and 0.3557 (right),
while in b) $\alpha=0.35739$ (left) and $\alpha=0.35745$ (right).
The dashed lines show the (near-)critical flow,
in a) $\alpha_c=0.3555842$ and in b) $\alpha_c\approx0.357992$.
The flows near criticality in a) and b) are qualitatively different.
No critical fixed-point structure emerges in b), instead the level spacing
decreases along the flow -- this is a signature of the flow towards a
Gaussian fixed point.
(The critical flow in b) cannot be followed beyond $n\sim 20$ due to
Hilbert-space truncation errors.)
}
\label{fig:flows04}
\end{figure}

\begin{figure}[!t]
\epsfxsize=3.6in
\centerline{\epsffile{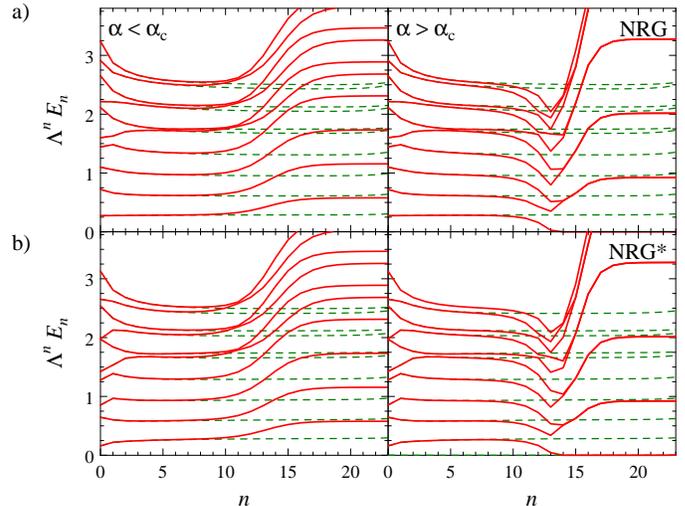}}
\caption{
As in Fig.~\ref{fig:flows04}, but for $s=0.6$ and $\kappa=0.8$ in b).
In a), $\alpha=0.6702$ (left), 0.6704 (right), $\alpha_c=0.6702965$ (dashed),
while in b) $\alpha=0.6723$ (left), $\alpha=0.6725$ (right), and
$\alpha_c=0.67240901$ (dashed).
Here, the flows in a) and b) are qualitatively similar,
with the level structure at the critical fixed point in b) slightly
deviating from that in a).
}
\label{fig:flows06}
\end{figure}

Doing so, we obtain the flow diagrams from the mass-flow corrected
NRG$^\ast$ algorithm, which represent a central result of this paper.
Those are shown in Figs.~\ref{fig:flows04} and \ref{fig:flows06} for
$s=0.4$ and 0.6, respectively, together with the flow diagrams from standard
NRG. The latter are similar to the ones shown in earlier papers.\cite{BTV,BLTV}

Let us start the discussion with Fig.~\ref{fig:flows04}a, displaying the standard NRG
flow for $s=0.4$ near the critical coupling strength.
For $\alpha\lesssim\alpha_c$ (left) the flow reaches the delocalized fixed point,
whereas it is directed towards the localized fixed point for $\alpha\gtrsim\alpha_c$ (right);
note that the latter is not correctly described due to Hilbert-space
truncation.\cite{BLTV}
The flow at $\alpha_c$ (dashed) shows a different NRG fixed point, which has been identified
with the critical fixed point. For both $\alpha\lesssim\alpha_c$ and $\alpha\gtrsim\alpha_c$
this level structure is visible at intermediate stages of the flow, before the system
departs towards one of the stable fixed points -- this crossover is usually identified
with the quantum critical crossover scale $T^\ast$ above which the system is critical.
Now consider the flow of NRG$^\ast$, Fig.~\ref{fig:flows04}b,
which includes the mass-flow correction of Sec.~\ref{sec:cure}.
While the asymptotic fixed points for both $\alpha<\alpha_c$ and
$\alpha>\alpha_c$ are identical, the flow near criticality is strikingly different.
In particular, no stable level pattern emerges, possibly corresponding to a critical NRG
fixed point. Instead, all levels appear to converge toward zero energy before the
critical regime is left.
Note that the critical flow cannot be followed to large $n$
(the system is always localized or delocalized for $n\gtrsim 20$).

In Fig.~\ref{fig:flows06}, the same comparison of flow diagrams is given for $s=0.6$.
Here, no qualitative difference between the flows without and with mass-flow
correction is seen. A stable level pattern is visible near criticality in both cases, but the
level energies differ slightly in Fig.~\ref{fig:flows06}a and b.
We found this behavior to be generic for $1/2<s<1$, while the absence of a critical NRG
fixed point as in Fig.~\ref{fig:flows04}b is characteristic for all $0<s<1/2$,
if $\kappa$ is chosen according to the criteria in Sec.~\ref{sec:cureint}.

It is straightforward to discuss what would be expected for a quantum phase transition
above its upper-critical dimension. The Gaussian fixed point features free massless bosons,
and interactions are required to stabilize the system at $T>0$.
Those are dangerously irrelevant and flow to zero in the critical regime, with a scaling
dimension which is small near the upper-critical dimension.
Translated into a many-body spectrum, this implies that {\em at} the Gaussian critical
fixed point the spectrum consists of an infinite number of degenerate levels at zero
energy, while the flow towards the critical fixed point is characterized by the level spacing
flowing to zero as $n\to\infty$.
The latter is precisely what is seen in Fig.~\ref{fig:flows04}b.
It is also clear that within NRG$^\ast$ the fixed point itself can never be reached,
because with decreasing interactions (i.e. decreasing level spacing) the error
introduced by the Hilbert-space truncation becomes more and more serious
(i.e. bosonic occupation numbers become large).
This implies that small values of $T^\ast$ cannot be reached
(as the system always flows to either the localized or delocalized phase
below some $T_{\rm min}^\ast$) which also limits the precision with which we can
determine $\alpha_c$.

A few remarks are in order:
(i) During the flow towards the Gaussian fixed point, Fig.~\ref{fig:flows04}b,
the rate of decrease in level spacing as function of $n$ depends strongly on $s$,
i.e., the level spacing decays faster with smaller $s$, qualitatively consistent
with the scaling dimension of the interaction $u$ being\cite{luijten} $(2s\!-\!1)$.
Correspondingly, the critical flows breaks down earlier for smaller $s$.
(ii) The value of the critical coupling $\alpha_c$ differs between NRG and NRG$^\ast$.
As discussed above, this is a result of spectrum truncation within NRG$^\ast$.
We have checked that the difference decreases with increasing $N_s$.
Further the difference is larger for smaller $s$, which follows from the mass-flow
error itself being larger for smaller $s$, see Fig.~\ref{fig:reg}b.

We conclude that the critical behavior of the spin-boson model for $s<1/2$ is Gaussian.
The stable critical fixed point in Fig.~\ref{fig:flows04}a is then an artifact of the
mass-flow error, where the system follows the thick solid trajectory in
Fig.~\ref{fig:pdflow}.
In contrast, for $s>1/2$ the critical theory of the spin-boson model is interacting.
These conclusions are supported by the analysis of $\chi(T)$,
see next subsection.

\subsection{Susceptibility}

We continue with NRG results for the order-parameter susceptibility
\begin{equation}
\chi = d\langle\sigma_z\rangle/d\epsilon
\end{equation}
of the spin-boson model.
We will focus on the power-law behavior $\chi(T)\propto T^{-x}$ in the quantum-critical
regime.

\begin{figure}[!t]
\epsfxsize=3.4in
\centerline{\epsffile{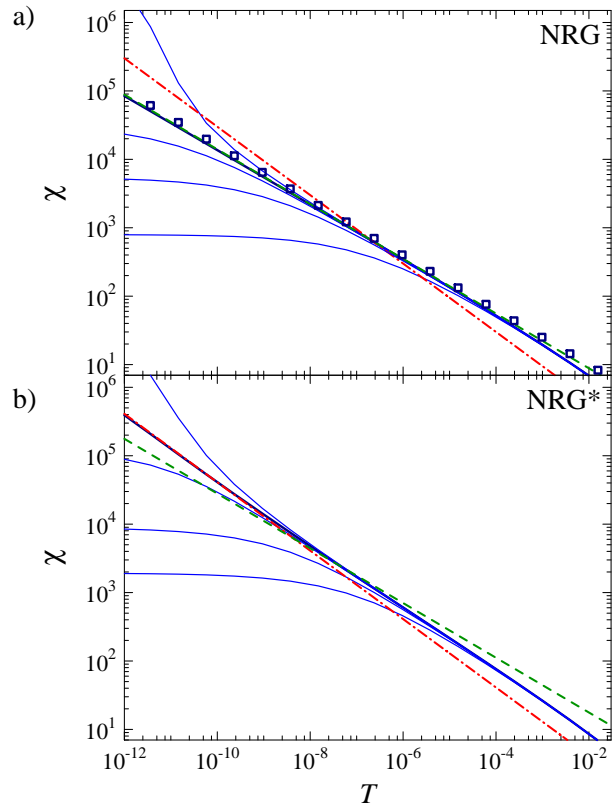}}
\caption{
Local susceptibility $\chi$ of the spin-boson model at $s=0.4$ obtained from
a) standard NRG without mass-flow correction and
b) NRG$^\ast$ with $\kappa=0.5$ mass-flow correction.
The $\alpha$ values are
a) 0.355, 0.3555, 0.35557, $0.3555842 = \alpha_c$, 0.3556,
b) 0.357, 0.3573, 0.35739, $0.3573992 \approx \alpha_c$, 0.35741,
with the critical (non-critical) $\chi$ plotted with thick (thin) lines.
The other parameters are as in Fig.~\ref{fig:flows04}.
The dashed (dash-dotted) lines show power laws with $T^{-s}$ ($T^{-1/2}$)
as reference.
The squares in a) show the $\chi$ of a harmonic oscillator,
calculated with a truncated Wilson chain corresponding to the $\alpha_c$
of NRG and $\Omega$ tuned to resonance.
The critical $\chi$ in panel b) is seen to approach mean-field behavior at low $T$,
$\chi \propto T^{-1/2}$.
}
\label{fig:chi04}
\end{figure}

\begin{figure}[!t]
\epsfxsize=3.4in
\centerline{\epsffile{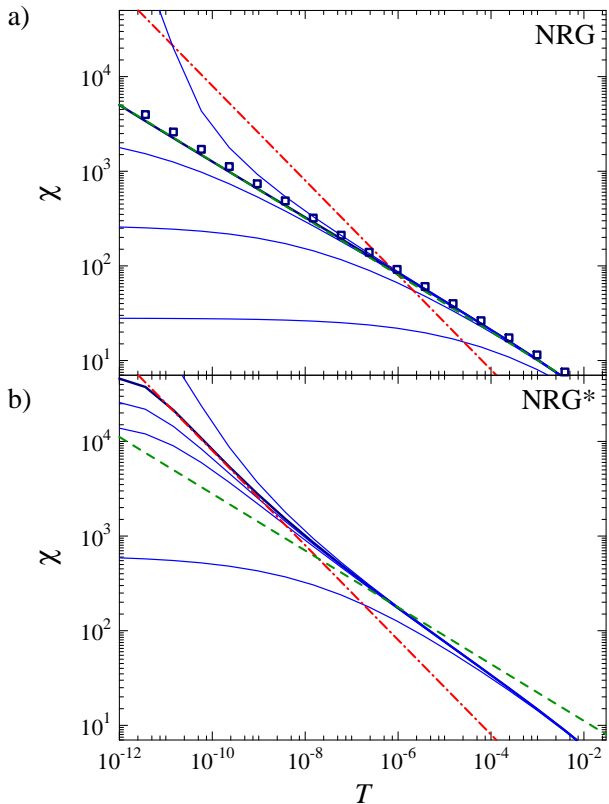}}
\caption{
Same as Fig.~\ref{fig:chi04}, but for $s=0.3$ and $\kappa=0.36$ in b).
The $\alpha$ values are
a) 0.23, 0.238, 0.2388, $0.238885 = \alpha_c$, 0.239,
b) 0.24, 0.2408, 0.24083, $0.24085 \approx \alpha_c$, 0.2409.
Note that $T_{\rm min}^\ast \approx 10^{-11}$, below which the near-critical
curves are affected by Hilbert-space truncation.
}
\label{fig:chi03}
\end{figure}

For both $s=0.4$ and $s=0.3$, data from both standard NRG and NRG$^\ast$ are shown in
Figs.~\ref{fig:chi04} and \ref{fig:chi03}, respectively.
As reported before, $x=s$ is obtained from NRG, Fig.~\ref{fig:chi04}a and Fig.~\ref{fig:chi03}a,
while the correct result near a Gaussian fixed point is $x=1/2$.
It should be noted that this $T^{-1/2}$ power law
requires the renormalized quartic interaction to be small.
However, once the effective interaction becomes small in the numerics,
the NRG$^\ast$ algorithm breaks down due to Hilbert-space truncation.
Thus, the weakly interacting Gaussian critical regime cannot be reached, and
we cannot expect to see an asymptotic $T^{-1/2}$ susceptibility power law.
For our parameter values, the truncation-induced lower cutoff scale, $T_{\rm min}^\ast$,
for the critical regime is $\mathcal{O}(10^{-13})$ for $s=0.4$ and $\mathcal{O}(10^{-11})$
for $s=0.3$.
Notably, the NRG$^\ast$ results in Figs.~\ref{fig:chi04}b and Fig.~\ref{fig:chi03}b
{\em do} follow $T^{-1/2}$ over two to three decades in temperature above $T_{\rm min}^\ast$,
while $T^{-s}$ is never seen.

We are again forced to conclude that the $T^{-s}$ behavior in standard NRG,
Figs.~\ref{fig:chi04}a and \ref{fig:chi03}a, is an artifact of the mass-flow error.
To support this, we also show the susceptibility
of the dissipative harmonic oscillator model, Eq.~\eqref{chistat}, calculated using a truncated
Wilson chain with the {\em same} chain parameters as in the NRG run for the spin-boson
model and $\Omega$ tuned to resonance.
As explained in Sec.~\ref{sec:dho}, this model has $\chi=\infty$, but a finite $\chi$
results exclusively from the mass-flow error.
Remarkably, this $\chi$ matches the $\chi$ from NRG for the spin-boson model
at low temperatures to an accuracy of better than 15\% -- this is consistent with the
assertion that the latter reflects the physics of a Gaussian fixed point
artificially stabilized by the mass-flow effect.

\begin{figure}[!t]
\epsfxsize=3.4in
\centerline{\epsffile{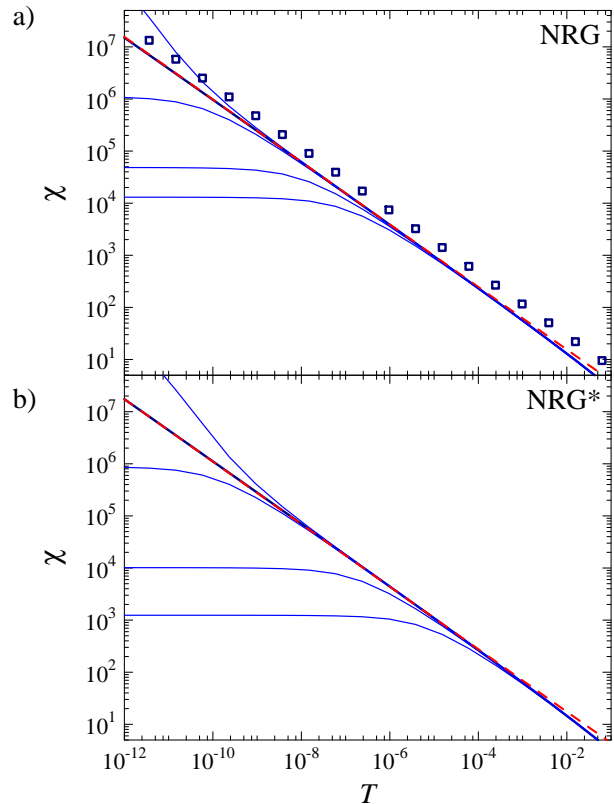}}
\caption{
Same as Fig.~\ref{fig:chi04}, but for $s=0.6$ and $\kappa=0.8$ in b).
The $\alpha$ values are
a) 0.67, 0.6702, 0.67029, $0.67029648 = \alpha_c$, 0.6703,
b) 0.67, 0.672, 0.6724,   $0.67240901 = \alpha_c$, 0.67242.
The dashed line shows a power law with $T^{-s}$.
}
\label{fig:chi06}
\end{figure}

Susceptibility data for $s=0.6$ are shown in Fig.~\ref{fig:chi06}.
Both NRG and NRG$^\ast$ yield a power law with $x=s$, albeit with prefactors differing
by 10\%.
Here, the harmonic-oscillator $\chi$ (with mass flow) and the NRG $\chi$
do {\em not} match, but instead differ by roughly a factor of 1.9.
All this is consistent with a true interacting fixed point.

\subsection{Other observables and exponents}

From the discussion, it is obvious that other observables {\em at} criticality will
suffer the mass-flow error similar to $\chi(T)$. This applies to thermodynamic
quantities including entropy and specific heat, but also to zero-temperature
dynamic quantities, like the susceptibility $\chi(\w)$.
While the latter is defined from the ground state, the corresponding NRG evaluation is in
fact done during the flow.\cite{nrgrev}
Hence, $\chi(\w)$ for $s<1/2$ is potentially incorrect as well.
However, $\chi(\w)$ can be proven to follow $\w^{-s}$ for all $s$,
irrespective of whether the fixed point is Gaussian or interacting,\cite{fisher,suzuki}
such that the mass-flow effect only introduces quantitative deviations.

Off-critical properties are to leading order {\em not} affected by the mass-flow error,
because the artificial mass vanishes as $T\to 0$ while the physical mass remains finite.
However, subleading corrections are subject to the mass-flow error.

The NRG calculations of Ref.~\onlinecite{VTB} did not only find the critical exponent $x$
to deviate from its mean-field value for $s<1/2$, but also the order-parameter exponents
$\beta$ and $\delta$.
As discussed in detail in Ref.~\onlinecite{erratum}, this incorrect result is due to a
{\em different} failure of the bosonic NRG, namely the fact that the Hilbert-space truncation
prevents an asymptotically correct representation of the localized fixed point for $s<1$.
For mean-field criticality, $\beta$ and $\delta$ are not properties of the critical fixed
point, but instead of the flow towards the localized fixed point.
As the latter suffers from the Hilbert-space truncation,
$\beta$ and $\delta$ are unreliable.
However, large values of $N_b$ can be used to uncover the physical power laws at
intermediate scales (which are of mean-field type for $s<1/2$),
before truncation effects set in.\cite{erratum}


\section{Other models}

For both versions of the anharmonic oscillator, Eqs.~\eqref{dao} and \eqref{dao2}, we
have obtained results which are qualitatively similar to those for the spin-boson model.
In particular, the standard NRG exhibits signatures of an interacting critical fixed
point for all $s$, as in Figs.~\ref{fig:flows04}a and \ref{fig:flows06}a.
For $\mathcal{H}'_{\rm DAO}$, this result is obviously incorrect for $s<1/2$,
due to its equivalence to a local $\phi^4$ theory.
Accordingly, the mass-flow corrected NRG$^\ast$ algorithm yields Gaussian behavior in
both models for $s<1/2$.
Hence, all models (\ref{dao},\ref{dao2},\ref{sbm}) belong to the same universality class
and follow the quantum-to-classical correspondence.

We have also investigated the mass-flow effect for particle-hole asymmetric fermionic
impurity models. While generically present, its effects on observables
turn out to be tiny, for details see App.~\ref{app:rlm}.

Finally, a remark on symmetries and the quantum-to-classical correspondence is in order:
While all cases discussed so far feature
Ising-symmetric critical degrees of freedom, impurity spin models with higher symmetry
[e.g. SU(2)] have been discussed extensively in the literature as well.
Here, a direct quantum-to-classical mapping (via a re-interpretation of the
Trotter-discretized action of the quantum model after integrating out the bath)
is usually not possible, due to the impurity spin's Berry phase.
Indeed, the so-called Bose-Kondo model with SU(2) symmetry exhibits a stable
intermediate-coupling fixed point\cite{sbv} (unlike any classical 1d spin model),
and the SU($N$)-symmetric Bose-Fermi Kondo model has been shown to display
a quantum critical point with hyperscaling for all $s$.\cite{kisi}


\section{Conclusions}

In this paper, we have investigated a source of error in Wilson's NRG method which had
received little attention before. This mass-flow error is inherent to the iterative
diagonalization scheme of NRG which neglects the low-energy part of the bath when
calculating observables.

We have traced the mass-flow effect in the dissipative harmonic oscillator model, where
results for the finite-temperature susceptibility turn out to be qualitatively incorrect
in general.
Applied to quantum phase transitions in bosonic impurity models, we have argued that the
mass-flow effect introduces qualitative errors in the critical regime of
mean-field quantum phase transitions,
while it only leads to quantitative errors for interacting quantum criticality.
A simple extension of the NRG algorithm allows to cure the mass-flow error
asymptotically near the fixed points of interest.
We have applied this modified algorithm to the sub-ohmic spin-boson model and found unambiguous
signatures of mean-field behavior for $s<1/2$,
including a flow towards a Gaussian critical fixed point, Fig.~\ref{fig:flows04}b,
and a susceptibility power law with mean-field exponent, Figs.~\ref{fig:chi04}b and \ref{fig:chi03}b.
We have thus resolved the discrepancy between results from NRG and those from
other numerical methods.\cite{VTB,rieger,werner,fehske}

As the conventional NRG is not capable of describing mean-field critical points,
claims of non-mean-field behavior in related\cite{bosonization_foot} Ising-symmetric
impurity models with sub-ohmic bosonic bath\cite{kevin,si09,glossop09} need to be
re-visited.


\acknowledgments

We thank S. Florens, E. G\"artner, K. Ingersent, S. Kirchner, R. Narayanan, H. Rieger,
Q. Si, and T. Voj\-ta for discussions.
This research was supported by the DFG through SFB 608 (MV,RB), FG 538 (MV), and FG 960 (MV,RB).
MV also acknowledges financial support by the Heinrich-Hertz-Stiftung NRW and the
hospitality of the Centro Atomico Bariloche where part of this work was performed.


\appendix

\section{Spin-boson model: Determining the proper mass-flow correction}
\label{app:kappa}

\begin{figure*}[!t]
\epsfxsize=6.7in
\centerline{\epsffile{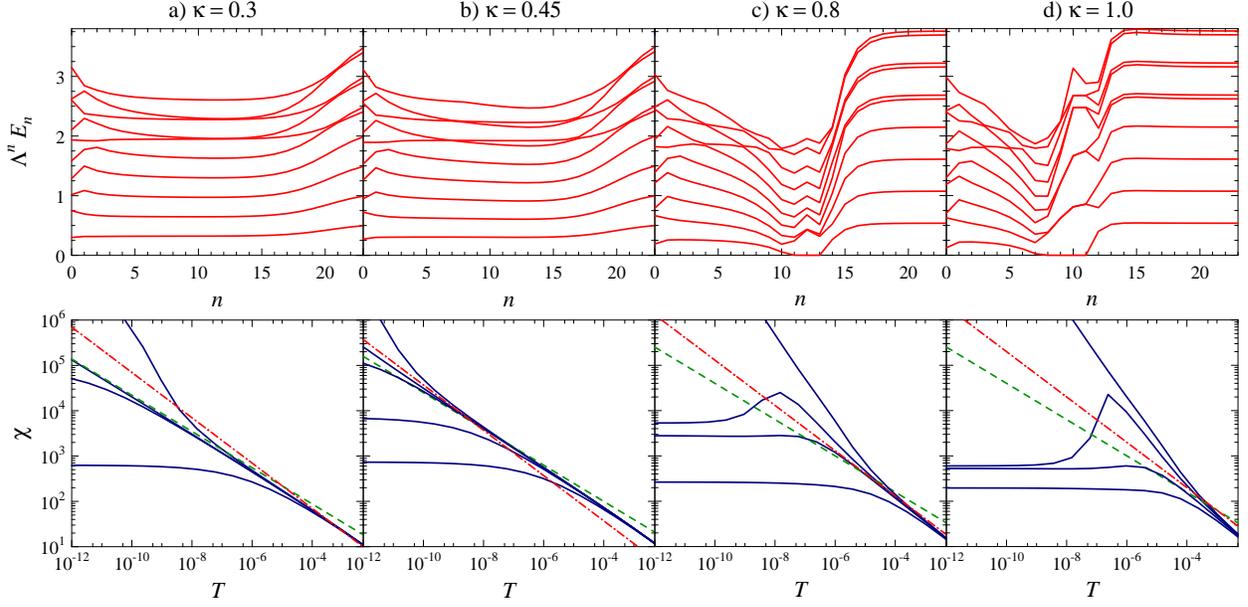}}
\caption{
NRG$^\ast$ results for the spin-boson model with
$s=0.4$, $\Lambda=4$, $N_b=12$, $N_s=40$,
for different values of the flow-correction parameter $\kappa$, Eq.~\eqref{kn2}.
Top:  Flow diagrams for $\alpha\lesssim\alpha_c$,
specifically $\alpha=0.3559$, 0.3569, 0.364, 0.372 from a) to d).
Bottom: Local susceptibilities $\chi$ for various $\alpha\sim\alpha_c$,
evaluated with $\bar\beta=1$.
The dashed (dash-dotted) lines show power laws with $T^{-0.4}$ ($T^{-1/2}$)
as reference. The $\alpha$ values are
a) 0.355, 0.3559, 0.3559076, 0.356,
b) 0.356, 0.3568, 0.356905, 0.3569096, 0.35692,
c) 0.36, 0.3635, 0.364, 0.365,
d) 0.365, 0.369, 0.372, 0.374.
Signatures of overcompensation, $\kappa>\kappa_0$, are obvious in c) and d),
while the critical power law $\chi\propto T^{-0.4}$ in a) implies undercompensation, $\kappa<\kappa_0$.
We conclude $\kappa_0 \approx 0.5$.
}
\label{fig:flows04kappa}
\end{figure*}

As discussed in Sec.~\ref{sec:cureint}, a general algorithmic solution to the mass-flow
problem for interacting bosonic impurity models is not available. Instead, we have argued that
an empirical correction via Eq.~\eqref{kn2} within the NRG$^\ast$ algorithm is appropriate,
with a prefactor $\kappa$ which depends on the fixed point of interest.

Here we show the influence of $\kappa$ on the NRG$^\ast$ results for the sub-ohmic
spin-boson model near criticality.
Fig.~\ref{fig:flows04kappa} displays NRG flows for $\alpha\lesssim\alpha_c$ (top)
and local susceptibility data for various $\alpha\sim\alpha_c$ (bottom)
for different values of $\kappa$ for $s=0.4$.
The central observation is that the behavior {\em qualitatively} changes when
$\kappa$ is varied from 0.3 to 1.0.

In a) both the critical and delocalized NRG fixed points are clearly visible in the flow,
and the critical $\chi(T)\propto T^{-s}$.
In b), the flow in the critical regime displays a decreasing level spacing with increasing
$n$, and no asymptotic $T^{-s}$ power law is observed.
Panels c) and d) show clear signs of overcompensation as discussed in Sec.~\ref{sec:cureint},
i.e., a non-monotonic flow (critical--localized--delocalized) and a corresponding
non-monotonic $\chi(T)$ near $\alpha_c$. Here, a precise determination of $\alpha_c$ is
impossible, and no critical power law in $\chi(T)$ emerges.
A detailed analysis of case a) shows that the behavior is qualitatively similar to that
of the standard NRG. For a Gaussian fixed point, this would imply undercompensation.
Together with the discussion in Sec.~\ref{sec:cureint}, these observations strongly suggest
that the critical fixed point of the spin-boson for $s=0.4$ is Gaussian,
with $\kappa_0 \approx 0.5$, see also the data in Fig.~\ref{fig:flows04}.
Indeed, the critical $\chi(T)$ in Fig.~\ref{fig:flows04kappa}b does not
follow $\chi\propto T^{-x}$ with $x=0.4$ at the lowest $T$ shown, but instead
crosses over to larger $x$.
A similar procedure for other $s<1/2$ yields
$\kappa_0(s\!=\!0.2) \approx 0.2$ and
$\kappa_0(s\!=\!0.3) \approx 0.35$.

\begin{figure}[!t]
\epsfxsize=3.6in
\centerline{\epsffile{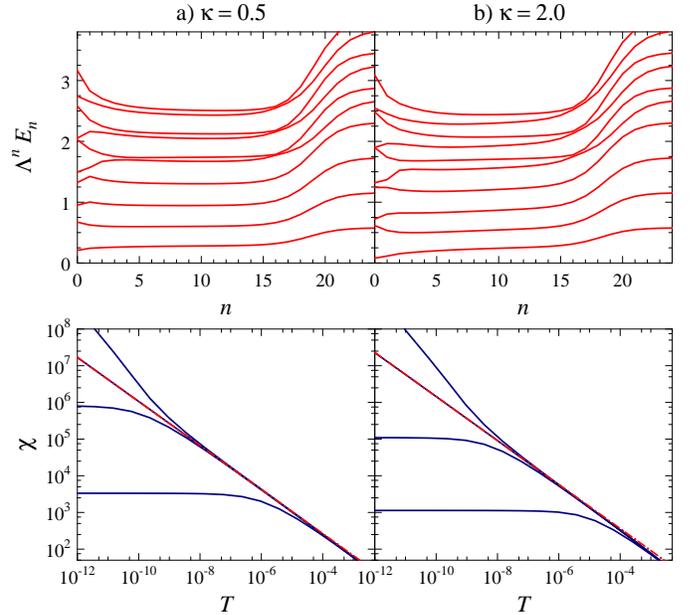}}
\caption{
As in Fig.~\ref{fig:flows04kappa}, but for $s=0.6$.
Here, no qualitative changes occur upon variation of $\kappa$.
The $\alpha$ values are
0.671007 and 0.693268 in the top panel, and
a) 0.67, 0.671, 0.67100924, 0.67102,
b) 0.69, 0.6932, 0.35739, 0.6932712, 0.6933
in the bottom panel.
The dashed line shows a power law with $T^{-0.6}$.
}
\label{fig:flows06kappa}
\end{figure}

In contrast, data for $s=0.6$, Fig.~\ref{fig:flows06kappa}, do {\em not}
display a qualitative change when $\kappa$ is varied from 0.5 to 2.0.
($\kappa=0$ data from the standard NRG are in Fig.~\ref{fig:flows06}.)
Instead, for all $\kappa$ a stable critical fixed-point spectrum emerges,
and the critical susceptibility follows $\chi\propto T^{-s}$.
This implies an interacting critical fixed point in the spin-boson model for $s=0.6$.
An extrapolation of $\kappa_0(s)$ suggests $\kappa_0(s\!=\!0.6)\approx0.8$
(which, however, is not very accurate, see Sec.~\ref{sec:cureint}).
We note that Figs.~\ref{fig:flows06kappa}a and b differ quantitatively:
The level structure at the critical fixed point is somewhat shifted,
and the prefactor of the critical power law of $\chi(T)$ is 40\% larger in b).
This trend simply reflects that larger $\kappa$ reduces the order-parameter
mass along the flow trajectory.


\section{Fermionic resonant-level model}
\label{app:rlm}

The mass-flow effect is in principle also present in fermionic impurity models
if the bath is particle-hole asymmetric in the low-energy limit.
Consider the spinless resonant-level model
\begin{eqnarray}
{\cal H}_{\rm RLM} &=&
\epsilon_f f^\dagger f +
\sum_i \lambda_i (f^\dagger c_i + {\rm h.c.}) +
\sum_i \w_i c_i^\dagger c_i\nonumber\\
\label{rlm}
\end{eqnarray}
with the bare level energy $\epsilon_f$. As in Sec.~\ref{sec:models}, one can define a
bath spectral density $J(\w)$, which, however, now generically has contributions at both
positive and negative frequencies.

The solution for the $f$ (impurity) Green's function is
\begin{equation}
G_f(\w) = \llangle f;f^\dagger \rrangle = \frac 1 {\w+i0^+ -\epsilon_f - \Gamma(\w)}
\end{equation}
The impurity properties of this model are non-singular except at resonance,
$\alpha=\alpha_c$, where $\epsilon_f + {\rm Re}\Gamma(\w\!=\!0) = 0$,
i.e., where the renormalized $f$ level coincides with the Fermi level.
The properties near resonance have been studied extensively in Refs.~\onlinecite{GBI,fritz04}
for the particle-hole symmetric case.

A mass-flow error arises only for bath spectra $J(\w)$ which are particle-hole asymmetric at
low energies. The low-energy asymmetry may be quantified by looking at
\begin{equation}
a(\w) = \frac {J(\w) - J(-\w)}{J(\w) + J(-\w)} \,.
\end{equation}
A finite $a(\w\!\to\!0)$ implies particle-hole asymmetry in leading order.
Otherwise the mass-flow error vanishes in the low-energy limit, this applies e.g. to a
metallic fermionic bath spectrum with different positive and negative band cutoff
energies.

We have studied the mass-flow error for the resonant-level model \eqref{rlm} with
a maximally particle-hole asymmetric power-law bath, i.e., $J(\w>0)\propto\w^s$ and $J(\w<0)=0$.
In analogy with Sec.~\ref{sec:dho}, we then expect a large mass-flow error at resonance.
Indeed, the resonance position is shifted in a similar
fashion as for the harmonic oscillator in Sec.~\ref{sec:dho}.
However, when comparing observables, like the $f$ level occupancy or its susceptibility,
calculated at $T_n$ for the semi-infinite and the truncated
chains with fixed $\epsilon_f$, we find that the differences are tiny (less than
$10^{-2}$), in stark contrast to the bosonic case.
The reason for the small mass-flow error is rooted in both the character of the observables
and the statistics of the particles. First, in the fermionic case all observables are
related to two-particle propagators, in contrast to the bosonic $\chi$ of Eq.~\eqref{chi_ho}.
Hence, the real part of $\Gamma(\w)$ never shows up as directly as in $\chi$,
due to a convolution integral.
Second, the response of fermions at resonance is less singular than that of bosons.
Therefore, deviations from the exact resonance condition have less consequences as
compared to the bosonic case.

In summary, the mass-flow error of NRG is present for particle-hole asymmetric fermionic
problems as well, but practically has little effect for the observables we have checked.



\begin{thebibliography}{99}

\bibitem{wilson75}
K. G. Wilson,
Rev. Mod. Phys. {\bf 47}, 773 (1975).

\bibitem{nrgrev}
R. Bulla, T. Costi, and T. Pruschke,
Rev. Mod. Phys. {\bf 80}, 395 (2008).

\bibitem{BTV}
R. Bulla, N. Tong, and M. Vojta,
Phys. Rev. Lett. {\bf 91}, 170601 (2003).

\bibitem{BLTV}
R. Bulla, H.-J. Lee, N. Tong, and M. Vojta,
Phys. Rev. B {\bf 71}, 045122 (2005).

\bibitem{kevin}
M. T. Glossop and K. Ingersent,
Phys. Rev. Lett. {\bf 95}, 067202 (2005);
Phys. Rev. B {\bf 75}, 104410 (2007).

\bibitem{mvrev}
M. Vojta, Phil. Mag. {\bf 86}, 1807 (2006).

\bibitem{bosonization_foot}
The Bose-Fermi Kondo model with a metallic bath density of states of fermions and an
Ising-symmetric bosonic bath is believed to have the same critical properties as the
spin-boson model, because the fermionic bath can be bosonized by standard techniques,
leading to a dissipative ohmic bath in addition to the Ising-symmetric bath. If the
latter has a sub-ohmic density of states, it dominates the critical properties, which
then are those of the sub-ohmic spin-boson model.
Results from NRG calculations\cite{kevin} are consistent with this expectation.

\bibitem{luijten}
E. Luijten and H. W. J. Bl\"ote,
\prb {\bf 56}, 8945 (1997).

\bibitem{fisher}
M. E. Fisher, S.-k. Ma, and B. G. Nickel,
\prl {\bf 29}, 917 (1972).

\bibitem{VTB}
M. Vojta, N. Tong, and R. Bulla,
Phys. Rev. Lett. {\bf 94}, 070604 (2005).

\bibitem{karyn}
K. Le Hur, P. Doucet-Beaupre, and W. Hofstetter,
Phys. Rev. Lett. {\bf 99}, 126801 (2007).

\bibitem{rieger}
A. Winter, H. Rieger, M. Vojta, and R. Bulla,
Phys. Rev. Lett. {\bf 102}, 030601 (2009).

\bibitem{werner}
P. Werner and M. Troyer,
unpublished.

\bibitem{fehske}
A. Alvermann and H. Fehske,
Phys. Rev. Lett. {\bf 102}, 150601 (2009).

\bibitem{erratum}
M. Vojta, N. Tong, and R. Bulla,
Phys. Rev. Lett. {\bf 102}, 294904(E) (2009).

\bibitem{si09}
S. Kirchner, Q. Si, and K. Ingersent,
Phys. Rev. Lett. {\bf 102}, 166405 (2009).

\bibitem{glossop09}
M. Cheng, M. T. Glossop, and K. Ingersent,
Phys. Rev. B {\bf 80}, 165113 (2009).

\bibitem{edmft}
  Q.~Si, S.~Rabello, K.~Ingersent, and J.~L.~Smith,
  Nature {\bf 413}, 804 (2001) and \prb {\bf 68}, 115103 (2003).

\bibitem{fritz04}
L. Fritz and M. Vojta,
Phys. Rev. B {\bf 70}, 214427 (2004).


\bibitem{GBI}
C.~Gon\-za\-lez-Buxton and K.~Ingersent,
Phys. Rev. B {\bf 57}, 14254 (1998).

\bibitem{weiss}
U. Weiss, {\it Quantum Dissipative Systems},
World Scientific (Singapore), 1993.

\bibitem{leggett}
A.~J. Leggett, S. Chakravarty, A.T. Dorsey, M.P.A. Fisher, A. Garg, and W. Zwerger,
Rev. Mod. Phys. {\bf 59}, 1 (1987).

\bibitem{KM96}
S. Kehrein and A. Mielke, Phys. Lett. A {\bf 219}, 313 (1996).

\bibitem{nrg_param}
The procedure of converting a given bath density of states into the parameters of the
Wilson chain is described in detail in Refs.~\onlinecite{nrgrev,BLTV}.

\bibitem{suzuki}
M. Suzuki, Prog. Theor. Phys. {\bf 49}, 424, 1106, 1440 (1973).

\bibitem{sbv}
S.~Sachdev, C. Buragohain and M.~Vojta,
Science {\bf 286}, 2479 (1999);
M.~Vojta, C.~Buragohain, and S.~Sachdev,
Phys. Rev. B {\bf 61}, 15152 (2000).

\bibitem{kisi}
L. Zhu, S. Kirchner, Q. Si, and A. Georges,
Phys. Rev. Lett. {\bf 93}, 267201 (2004);
S. Kirchner and Q. Si,
preprint arXiv:0808.2647.

\end{thebibliography}
\end{document}